\documentclass[nofootinbib,
               reprint,
               superscriptaddress,
               amsmath,amssymb,
               aps,
               float]{revtex4-2}

\usepackage[labelfont=bf,
   justification=Justified,
   format=plain]{caption}
\usepackage{gensymb}
\usepackage{graphicx}
\usepackage{bm}
\usepackage{hyperref}
\usepackage{siunitx}
\DeclareSIUnit\gauss{G}
\DeclareSIUnit\sccm{sccm}
\usepackage[bottom]{footmisc}

\begin{document}

\author{Zhen Han}
\email{zhenhan@uchicago.edu}
\affiliation{Department of Physics, University of Chicago, Chicago, IL 60637, USA}

\author{Zack Lasner}
\affiliation{Department of Physics, Harvard University, Cambridge, MA 02138, USA}
\affiliation{Harvard-MIT Center for Ultracold Atoms, Cambridge, MA 02138, USA}

\author{Collin Diver}
\affiliation{Center for Fundamental Physics, Northwestern University, Evanston, IL 60208, USA}

\author{Peiran Hu}
\affiliation{Department of Physics, University of Chicago, Chicago, IL 60637, USA}

\author{Takahiko Masuda}
\affiliation{Research Institute for Interdisciplinary Science, Okayama University, Okayama, 700-8530, Japan}

\author{Xing Wu}
\affiliation{Facility for Rare Isotope Beams, Michigan State University, East Lansing, MI 48824, USA}

\author{Ayami Hiramoto}
\affiliation{Department of Physics, Harvard University, Cambridge, MA 02138, USA}
\affiliation{Center for Fundamental Physics, Northwestern University, Evanston, IL 60208, USA}
\affiliation{Research Institute for Interdisciplinary Science, Okayama University, Okayama, 700-8530, Japan}
\affiliation{Harvard-MIT Center for Ultracold Atoms, Cambridge, MA 02138, USA}

\author{Maya Watts}
\affiliation{Center for Fundamental Physics, Northwestern University, Evanston, IL 60208, USA}

\author{Satoshi Uetake}
\affiliation{Research Institute for Interdisciplinary Science, Okayama University, Okayama, 700-8530, Japan}

\author{Koji Yoshimura}
\affiliation{Research Institute for Interdisciplinary Science, Okayama University, Okayama, 700-8530, Japan}

\author{Xing Fan}
\affiliation{Center for Fundamental Physics, Northwestern University, Evanston, IL 60208, USA}
\affiliation{Department of Physics, Harvard University, Cambridge, MA 02138, USA}
\affiliation{Harvard-MIT Center for Ultracold Atoms, Cambridge, MA 02138, USA}

\author{Gerald Gabrielse}
\affiliation{Center for Fundamental Physics, Northwestern University, Evanston, IL 60208, USA}

\author{John M. Doyle}
\affiliation{Department of Physics, Harvard University, Cambridge, MA 02138, USA}
\affiliation{Harvard-MIT Center for Ultracold Atoms, Cambridge, MA 02138, USA}

\author{David DeMille}
\affiliation{Department of Physics, University of Chicago, Chicago, IL 60637, USA}

\title{A cryogenic buffer gas beam source with in-situ ablation target replacement}

\date{\today}

\begin{abstract}
The design and performance of a Cryogenic Buffer Gas Beam (CBGB) source with a load-lock system is presented. The ACME III electron EDM search experiment uses this source to produce a beam of cold, slow thorium monoxide (ThO) molecules. A novel feature of the apparatus is the capability of replacing the ablation targets without interrupting the vacuum or cryogenic conditions, thus increasing the average signal in the eEDM search. The beam source produces $1.3\times10^{11}$ ground-state thorium monoxide (ThO) molecules per pulse on average, with rotational temperature of \SI{4.8}{\kelvin}, molecular beam solid angle of \SI{0.31}{\steradian}, and forward velocity of \SI{200}{\meter\per\second}, parameters that are consistent with the performance of a traditional source (without a load-lock) requiring time-consuming thermal cycles for target replacement. Long-term yield improvement of $\sim40\%$ is achieved when the load-lock system is employed to replace targets every two weeks.
\end{abstract}

\maketitle

\section{Introduction}

Cryogenic buffer gas beam (CBGB) sources produce atomic and molecular beams with high flux and low forward velocity \cite{hutzler2012buffer}. They are widely used in many experiments, including precision measurements \cite{grasdijk2021centrex, nl2018measuring, fitch2020methods,hutzler2020polyatomic}, cooling and trapping of atoms and molecules \cite{anderegg2018laser,mccarron2018magnetic,lasner2025magneto, anderegg2017radio, langin2023toward}, cold collisions \cite{wu2017cryofuge, kozyryev2015collisional}, and spectroscopy of clusters and organic molecules \cite{patterson2013enantiomer,tokunaga2017high,straatsma2017production}. In a CBGB, the atom or molecule of interest thermalizes with cryogenic buffer gas in the cell and is cooled in both translational and internal degrees of freedom, before exiting the cell and forming a directed molecular beam \cite{hutzler2012buffer}. The brightness of the resulting beam is determined by two main factors: the efficiency of introducing the species of interest into the buffer gas cell, and the efficiency of extracting it into a directed beam.

\begin{figure}
  \centering
  \includegraphics[width=\columnwidth]{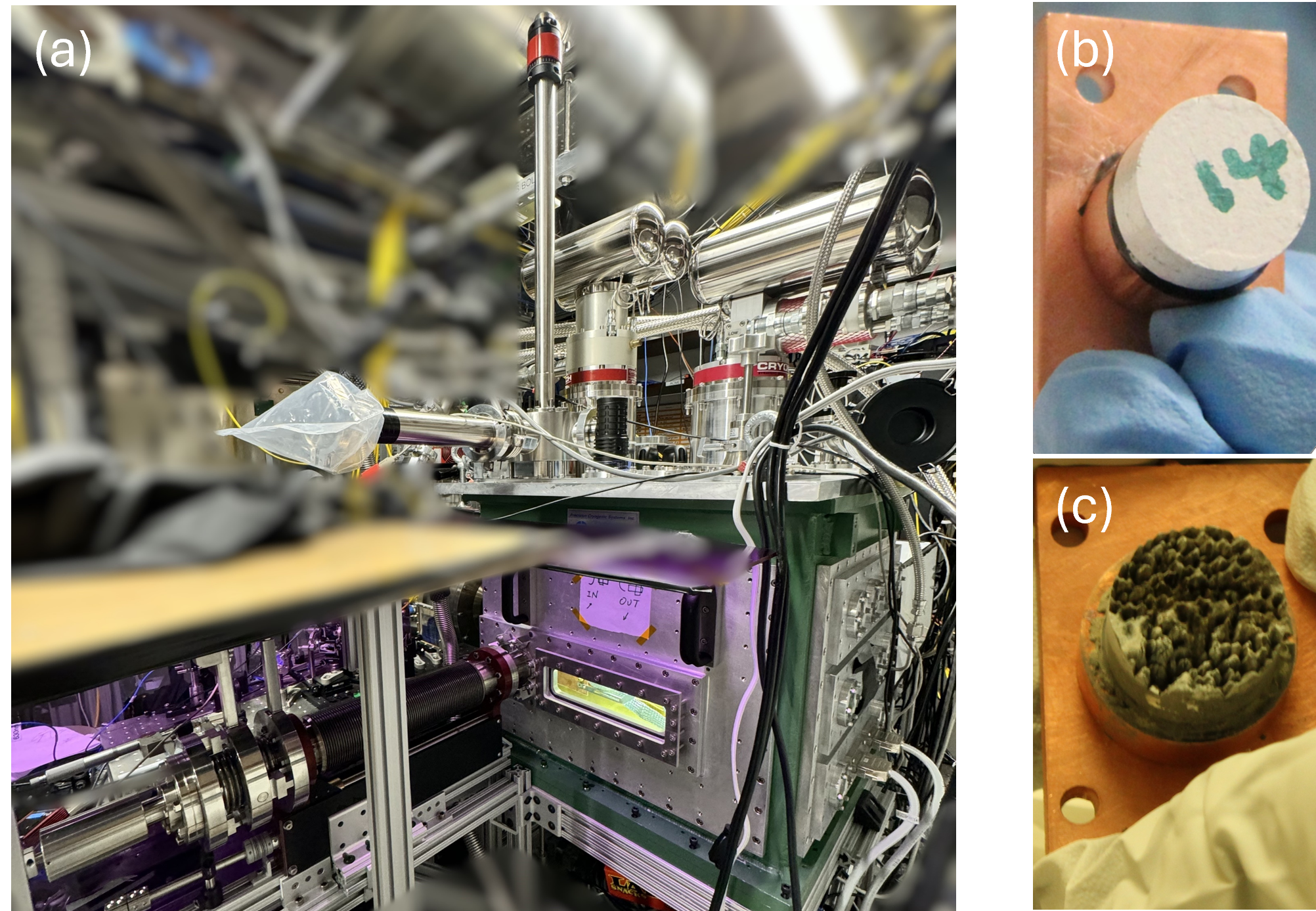}
  \caption{(a) ACME III CBGB with load-lock system. (b) Unused ThO$_{2}$ ablation target. (c) ThO$_{2}$ ablation target after $\sim30$ days of continuous use.}
  \label{fig:acutal_ll}
\end{figure}

Laser ablation is an effective method for introducing species of interest into the buffer gas cell, and has been successfully applied to produce many beams including SrF \cite{barry2011bright}, CaH \cite{vazquez2022direct}, BaF \cite{mooij2024influence}, YbOH \cite{jadbabaie2020enhanced}, MgF, Al, Yb \cite{wright2023cryogenic}. Typically a high-energy ($\sim$5--\SI{50}{\milli\joule}) pulsed laser is focused onto a solid precursor target. Desorption \cite{brewer2004laser}, vaporization \cite{maxwell2005high}, or chemical reactions with a reagent gas \cite{jadbabaie2020enhanced}, or within the ablation plume \cite{hutzler2011cryogenic}, produce the species that will be used in the downstream experiment. In many cases, the molecular yield of the ablation process is affected by the degradation of the surface condition of the ablation target, leading to a gradual decline in yield by repeated ablation shots \cite{hutzler2011cryogenic,barry2011bright}. In the previous ACME II electron electric dipole moment experiment \cite{acme2018improved}, where ThO was produced by ablation of ThO$_2$ ceramic targets\footnote{Produced by Idaho National Laboratory.}, a few days of continuous ablating would deplete the ``fresh surface'' of the ThO$_2$ target, leaving a porous and uneven surface and reduced ThO yield (Fig.\ref{fig:acutal_ll}~(b),~(c)). Replacement of the ablation target can maintain a high stable yield, but the replacement process introduces significant challenges. In a conventional CBGB, replacing the target requires venting the vacuum chamber to air, and the cryogenic components inside have to undergo a full warm-up and cool-down cycle, interrupting data collection for $\sim$ 1 day. This downtime is even longer when working with radioactive materials, such as ThO$_2$, because additional safety precautions and handling protocols extend the turnaround time. In-situ replacement of the ablation target without breaking vacuum or warming up can improve the long-term average yield. This is especially important in precision measurement experiments with long and continuous data taking.

\begin{figure*}
\centering
    \begin{minipage}{7in}
        \centering
    \includegraphics[width=\textwidth]{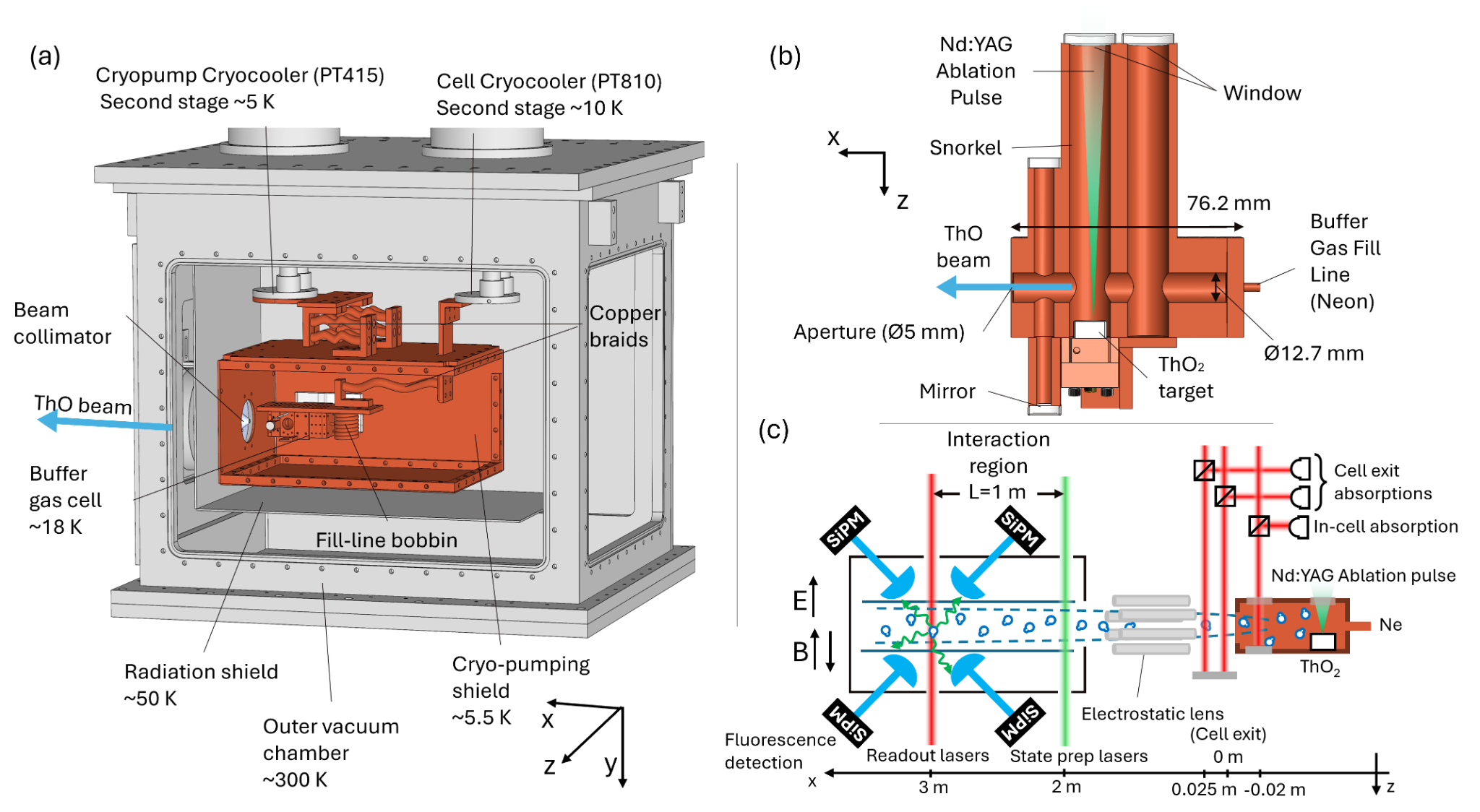}
    \end{minipage}
	    \caption{(a) Diagram of the ACME III beam box. (b) Section through the buffer gas cell. (c) Setup for characterizing beam properties using absorption spectroscopy and precession phase measurement.}
    \label{beambox}
\end{figure*}

Here we present the design and first implementation of a novel load-lock system for the ACME III CBGB (Fig.~\ref{fig:acutal_ll}(a)), which enables in-situ replacement of ThO$_2$ ablation targets without interrupting vacuum or cryogenic conditions---a capability not available in previous CBGB systems. We present CBGB ThO molecular yield, cell extraction fraction \cite{takahashi2021simulation}, rotational temperature, molecular beam divergence and forward velocity. We measure the long-term usable ThO molecular yield improvement realized by regularly replacing the ablation target approximately every two weeks.

\section{Experiment Setup}

\subsection{Cryogenic Buffer Gas Beam Source}
The general design, construction and performance of CBGB is described in ref.\cite{hutzler2012buffer}. Every CBGB in a given experiment is tuned to the molecule or atom of interest. In the case of ACME, neon buffer gas is used and ThO is the molecule.
The CBGB source used for the production of ThO molecules in the ACME III experiment is illustrated in Fig.\ref{beambox}~(a). ThO molecules are generated inside a buffer gas cell via laser ablation of a ThO$_2$ ceramic target with a pulsed Nd:YAG laser. Following ablation, the ThO molecules thermalize through collisions with the $\sim$\SI{18}{\kelvin} neon buffer gas before exiting the cell through the cell aperture (\SI{5}{\milli\meter} diameter). As the molecular beam propagates, most of the neon is scattered by the beam collimator and is cryo-pumped by the \SI{4}{\kelvin} shields (the ``cryopumping shields''). The \SI{50}{\kelvin} radiation shields also minimize the thermal load from black-body radiation on the cryo-pumping shields.

The buffer gas cell is made of copper\footnote{C10100 Oxygen-Free Electronic (OFE) Copper.\label{copper}} and has a cylindrical inner bore with a diameter of \SI{12.7}{\milli\meter} and a length of \SI{76.2}{\milli\meter} (Fig.\ref{beambox}~(b)). The neon buffer gas is supplied with a fill line running from the external gas handling system, through holes in shields, wound onto a bobbin for thermalization to the cell temperature, and then connected to the rear of the cell. The ThO$_2$ ceramic ablation target is \SI{13}{\milli\meter} diameter, \SI{9}{\milli\meter} thick cylinder. A Nd:YAG laser pulse with wavelength of \SI{1064}{\nano\meter}, pulse energy of \SI{20}{\milli\joule}, pulse duration of less than \SI{10}{\nano\second} is used with a repetition rate of \SI{50}{\hertz}. The Nd:YAG laser beam is focused to a diameter of less than \SI{50}{\micro\meter} on the target surface. To mitigate the coating of ablation debris on the ablation window, a `snorkel' places the ablation window $\sim$\SI{8}{\centi\meter} away from the ThO$_{2}$ target. A mirror positioned near the cell exit aperture retroreflects the probe laser beam used for in-cell ThO absorption spectroscopy.

\begin{figure*}
\centering
    \begin{minipage}{7in}
        \centering
    \includegraphics[width=\textwidth]{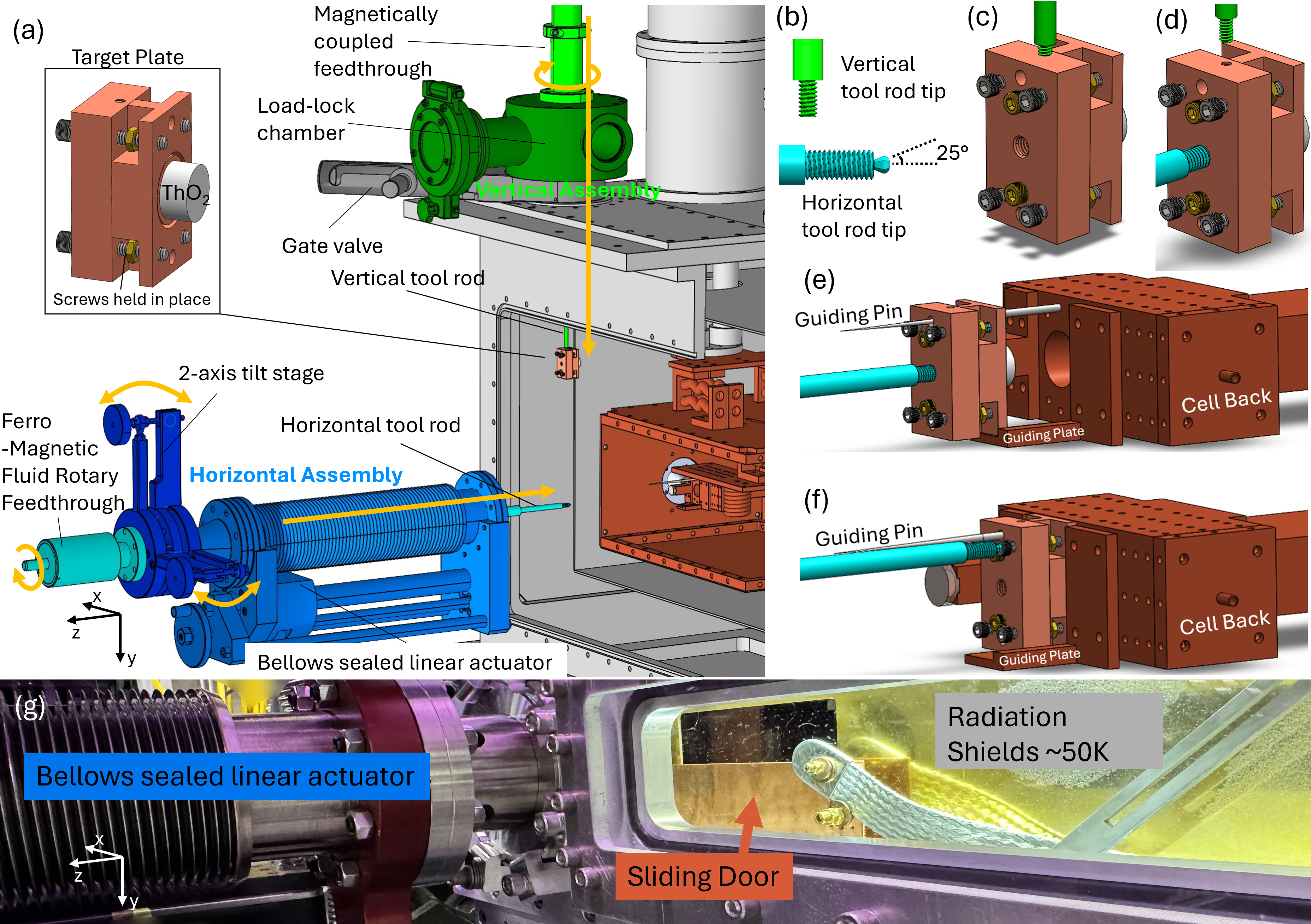}
    \end{minipage}
	    \caption{(a) Diagram of the load-lock system. The side panel of the beam box vacuum chamber, radiation shield and cryo-pumping shield on the side of load-lock system are not shown. Manipulators in the vertical assembly, including the magnetically coupled feedthrough and vertical tool rod (light green), transfer the target from the load-lock chamber (dark green) into the CBGB without breaking the vacuum and cryogenic environment. The horizontal assembly, which consists of a ferro-magnetic fluid rotary feedthrough and horizontal tool rod (cyan), 2-axis tilt stage (dark blue) and bellows sealed linear actuator (blue), serves to install the target into the cell. (b) Tips for vertical tool rod (light green) and horizontal tool rod (cyan). (c)-(f) Steps of transferring and installing the target plate onto the buffer gas cell. (g) Sliding door for blocking black-body radiation onto the cryo-pumping shields.}
    \label{loadlock}
\end{figure*}

The cryo-pumping shields, radiation shields, and the cell are structurally supported from the top or bottom beambox plates using thermally insulating materials such as stainless steel and G10, minimizing heat conduction while maintaining mechanical stability. The cell and the shields are thermally connected to cryocooler stages via flexible copper braids\footnote{Cooner Wire NER 7710836 OFE, welded to copper lugs.} (Fig.\ref{beambox}~(a)), which help isolate mechanical vibrations from cryocooler stages. The cryo-pumping shields are thermally connected to the second stage of the cryopump cryocooler\footnote{\href{https://bluefors.com/products/pulse-tube-cryocoolers/pt415-pulse-tube-cryocooler/}{Cryomech PT415}.} and held at \SI{5}{\kelvin}. The buffer gas cell is thermally isolated from the cryo-pumping shields and connected to the second stage of the cell cryocooler\footnote{\href{https://bluefors.com/products/pulse-tube-cryocoolers/pt810-pulse-tube-cryocooler/}{Cryomech PT810}.} which has a nominal running temperature of \SI{10}{\kelvin}. During experimental operation, the cell temperature is maintained at \SI{18}{\kelvin} with heaters to prevent neon freezing. The radiation shields are cooled to \SI{50}{\kelvin} by the first cooling stages of both cryocoolers. The entire assembly is inside of a room-temperature vacuum chamber, which is maintained at pressures below \SI{6e-8}{\milli\bar} during operation.

\subsection{Load‐Lock System}

The load-lock system, shown in Fig.\ref{loadlock}~(a), enables the replacement of the ablation target inside of the cell without interrupting the vacuum or cryogenic conditions of the buffer gas beam source. It consists of two additional assemblies that work in tandem, both mounted onto the beam source chamber. The vertical assembly attached to the top plate of the beam box transfers the target plate into the beam box under vacuum; the horizontal assembly attached to the side of the beam box manipulates the target plate and installs it into the cell. 

The ablation target is bonded using epoxy\footnote{Stycast 2850FT, with Loctite CAT 24LV catalyst.} to the target plate, which is transferred through the load-lock system. The screws for securing the target plate to the cell are held in place using nuts (Fig.\ref{loadlock}~(a)) and transferred along with the target plate. The target plate is made of copper\footref{copper}. For other types of ablation targets different bonding methods might be required in case of a larger thermal expansion coefficient mismatch between the target and target plate.

The vertical assembly includes a magnetically coupled feedthrough\footnote{\href{https://thermionics.com/uhv-products/multi-motion-feedthroughs-fllre-275-38/}{Thermionics FLLRE-275-38-21/CL/TM}.} that provides the vertical tool rod with linear motion to transfer the target plate, as well as rotational motion to thread onto the target plate. A load-lock chamber, pumped by a small turbo pump, connects to the beam box through a gate valve.

The horizontal assembly contains a ferro‐magnetic fluid rotary feedthrough\footnote{\href{https://seals.ferrotec.com/products/ferrofluidic-vacuum-rotary-feedthroughs/103909/}{Ferrotec SS-500-SLCB}.} on a 2-axis tilt stage\footnote{\href{https://thermionics.com/uhv-products/gblm-1-5-precision-tilt-stage/}{Thermionics TLTM-2.5-B450C-T450C-2}.} to tighten all screws to mount the target plate, and a bellows sealed linear actuator\footnote{\href{https://thermionics.com/uhv-products/z-series/}{Thermionics Z-B450C-T450T-2.50-10/SB}.} to position the target plate in front of the cell. The horizontal tool rod can hold the target plate via the threads near its end, and also has a ball-end hex driver tip with a \SI{25}{\degree} access angle (Fig.\ref{loadlock}~(b)) that is used to tighten the screws that attach the target plate to the cell, despite slight misalignment.

To install the target plate onto the cell under vacuum and cryogenic conditions, the target plate is first mounted on the vertical tool rod inside the load‐lock chamber (Fig.\ref{loadlock}~(c)). After pumping the load-lock chamber down with a turbo pump, the gate valve between the load-lock chamber and the beam box is opened, and the target plate is lowered to the location of the horizontal tool rod. The horizontal tool rod is then rotated to thread onto the target plate before the vertical rod is rotated and detached (Fig.\ref{loadlock}~(d)). Next, the bellows sealed linear actuator translates the target plate toward the cell, aligning it via a guiding pin and a guiding plate on the cell (Fig.\ref{loadlock}~(e)). Once the target plate is in position, the horizontal tool rod is detached and moved to the locations of mounting screws using the 2-axis tilt stage (Fig.\ref{loadlock}~(f)). Finally all mounting screws are tightened by employing the ferro‐magnetic fluid rotary feedthrough, completing the target installation. 

There are rectangular holes of dimension \SI{5}{\centi\meter} $\times$ \SI{8}{\centi\meter} on both the cryo-pumping shield and the radiation shield on the load-lock side, providing both mechanical and visual access for the installation procedure (Fig.\ref{loadlock}~(g)). A sliding door thermally anchored to the radiation shields by flexible copper braids and attached to a linear motion feedthrough can be closed to block the potential black-body heat load (\(\sim\)\SI{2}{\watt}) through the rectangular holes. By closing this door the temperature of the cryo-pumping shields is reduced from \SI{6}{\kelvin} to \SI{5}{\kelvin}.

To minimize leakage of buffer gas from the cell into the beam box, indium gaskets are used between all cell components. However, in order to simplify the design of the target plate and the operation of the load-lock system, no gasket or indium seal is placed at the interface between the target plate and the cell. We believe this will not result in significant leakage, with the following reasoning: to ensure that the leakage flow rate remains negligible compared to the forward flow rate, the leak area $A_{\text{leak}}$ must satisfy $A_{\text{leak}} \ll A_{\text{aperture}}$, where $A_{\text{aperture}} \approx \SI{20}{\milli\meter\squared}$ is the size of the cell exit aperture. For a potential leak at the interface between the target plate and the cell, this condition is met if the gap size is smaller than \SI{0.1}{\milli\meter}, which is achievable with precision-machined components.

\subsection{Detection Setup}

We use absorption spectroscopy to detect ThO molecules at three different locations: inside of the buffer gas cell, immediately outside of the cell aperture ($\sim$\SI{2}{\milli\meter} downstream), and \SI{2.5}{\centi\meter} downstream from the cell aperture (Fig.\ref{beambox}~(c)). At each location, a probe laser addresses the $C(v^{\prime}=0)\leftarrow X^{1}\Sigma^{+}(v=0)$ transition of ThO. The probe laser is retroreflected inside of the beam box, and its transmitted power is recorded by photodiodes. The optical density of the beam, $OD$, is defined as $OD=-\ln(P/P_0)$, where $P$ and $P_0$ denote the transmitted laser power in the presence and absence of ThO molecules, respectively.

The forward velocity of the molecular beam is determined via spin precession measurements (Fig.\ref{beambox}~(c)) as follows. 
After leaving the cell and being focused by electrostatic lens \cite{wu2022electrostatic}, at $\sim$\SI{2}{\metre} downstream from the cell, the molecules are prepared into the $H^3\Delta_1(v=0,J=1)$ state with state preparation lasers. As the molecules traverse the interaction region, their spins precess in the presence of parallel electric (E) and magnetic (B) fields. The precession phase is then measured using a \SI{703}{\nano\meter} readout laser that drives the $I(v^\prime=0,J^\prime=1)\leftarrow H^3\Delta_1 (v=0,J=1)$ transition with rapidly switching polarization \cite{kirilov2013shot}, and the fluorescence signal from the decay of the $I$ state to the $X$ state is detected by SiPMs \cite{masuda2023high}. The Zeeman phase is extracted from the precession phase measured with opposite B field directions as
$\phi^{\mathcal{B}} = (\phi(\hat{B} \cdot \hat{z} = +1) - \phi(\hat{B} \cdot \hat{z} = -1))/2$.
The molecular forward velocity $v$ is then inferred from the relation $ \phi^{\mathcal{B}} = -g \mu_B |B| L/v$, where $|B|=$ \SI{100}{\micro\gauss} is the magnitude of the applied magnetic field, $L=$ \SI{1}{\metre} is the separation between the state preparation and readout lasers, and $g=-0.00440(5)$ is the $g$-factor for the $H^3\Delta_1(v=0,J=1)$ state \cite{petrov2014zeeman}.

\section{Measurement Results}
\subsection{Cryogenic Temperatures}

\begin{figure}[htbp]
  \centering
  \includegraphics[width=\columnwidth]{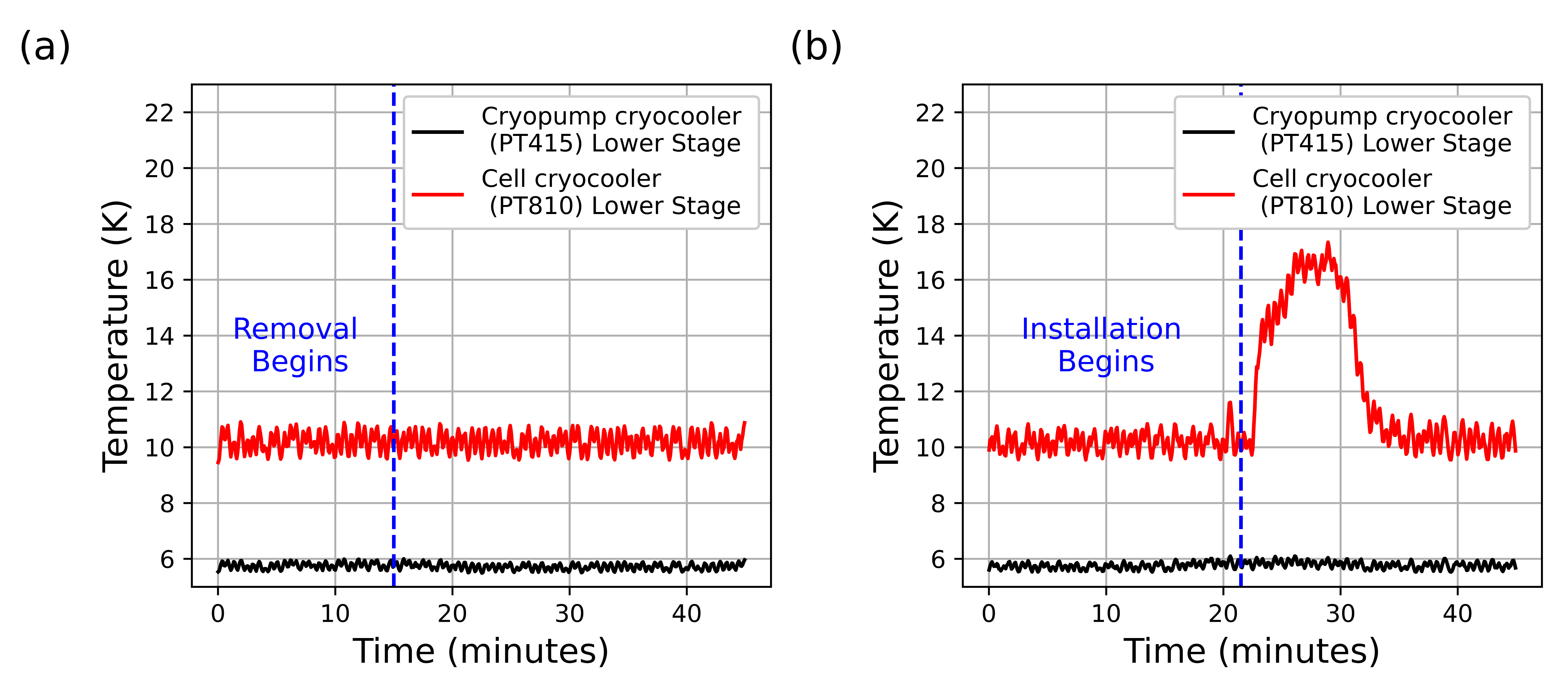}
  \caption{Temperatures of the cryogenic system during (a) removal of the target plate from the cell using the load lock system (b) installation of a room-temperature target plate directly onto the cell at cryogenic temperature using the load lock system. The cryopump cryocooler second stage is thermally connected to the \SI{4}{\kelvin} cryopumping shields, and the cell cryocooler second stage is connected to the buffer gas cell. Temperatures are measured using silicon diode sensors mounted on the cryocooler stages.}
  \label{fig:BBLLtemp}
\end{figure}

We replace the target assembly by first removing the old target and target plate and then installing a room-temperature replacement using the load-lock system, all while maintaining the cryogenic and vacuum conditions of the CBGB. Both removal and installation processes take \(\sim\)\SI{20}{\minute} to complete. The temperatures of the cryogenic system during target change are shown in Fig.~\ref{fig:BBLLtemp}.

\begin{figure*}
  \includegraphics[width=\textwidth]{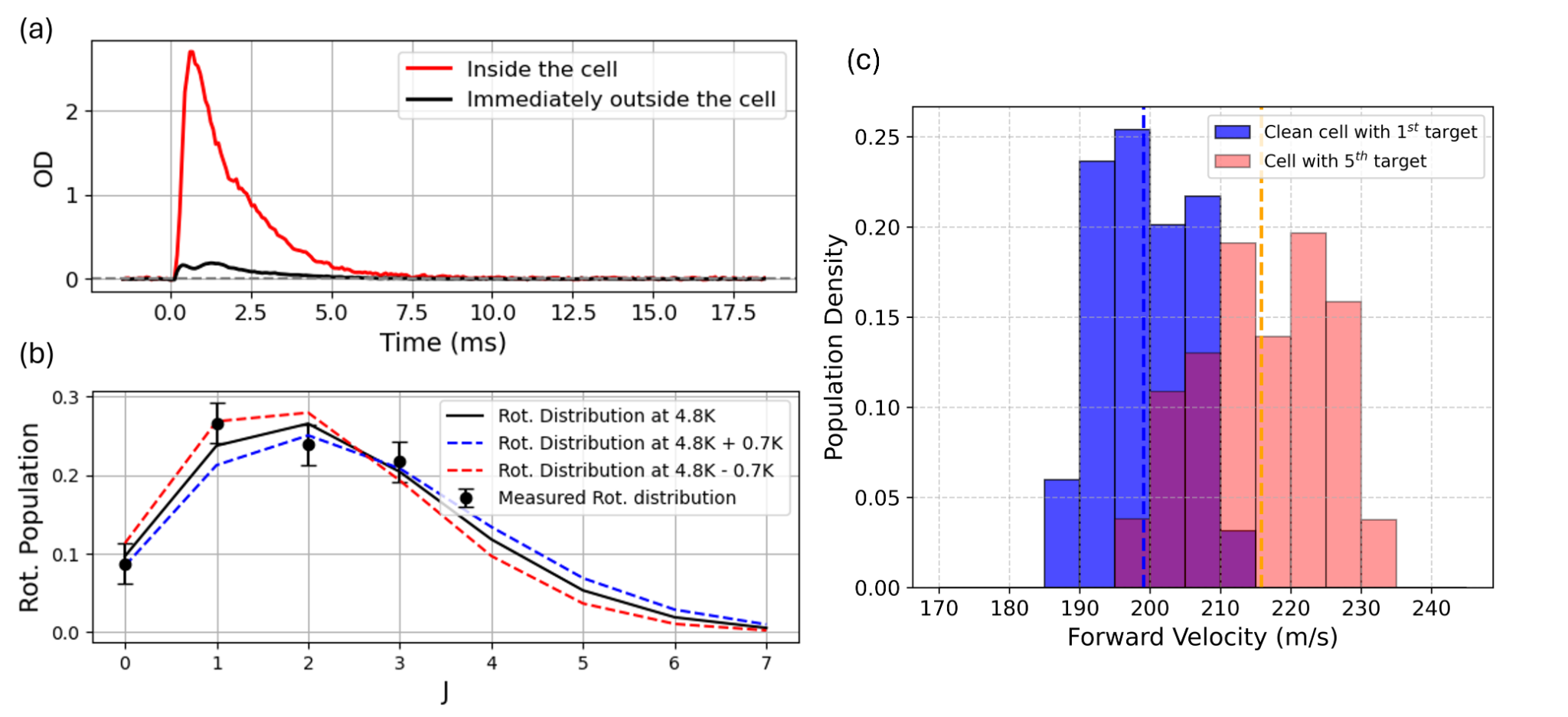}
  \caption{(a) A typical absorption measurement at a flow rate of \SI{40}{\sccm}. (b) Population distribution for different rotational levels. (c) Forward velocity distribution measured with clean cell and dusty cell. The vertical dashed lines represent the mean values of the distributions.}
  \label{fig:beam_properties_combined}
\end{figure*}

During normal operation of the beam source, the feedback-controlled heater on the cell continuously outputs $\sim$\SI{7}{\watt} to maintain the cell temperature at \SI{18}{\kelvin}, meaning that the system has at least $\sim$\SI{7}{\watt} of available cooling power when the room-temperature target plate (with total heat load of \(\sim\)\SI{3000}{\joule}) is installed.

During target assembly removal, the temperatures of the cryogenic components remain largely unchanged, primarily because of the good thermal insulation of the stainless steel horizontal tool rod that holds the target assembly. During installation, when the target plate at room temperature approaches the cell, the temperature of the cell cryocooler second stage temporarily increases before stabilizing at its baseline of \SI{10}{\kelvin} in $\sim$\SI{20}{\minute}.

\begin{figure*}
  \centering
  \includegraphics[width=\textwidth]{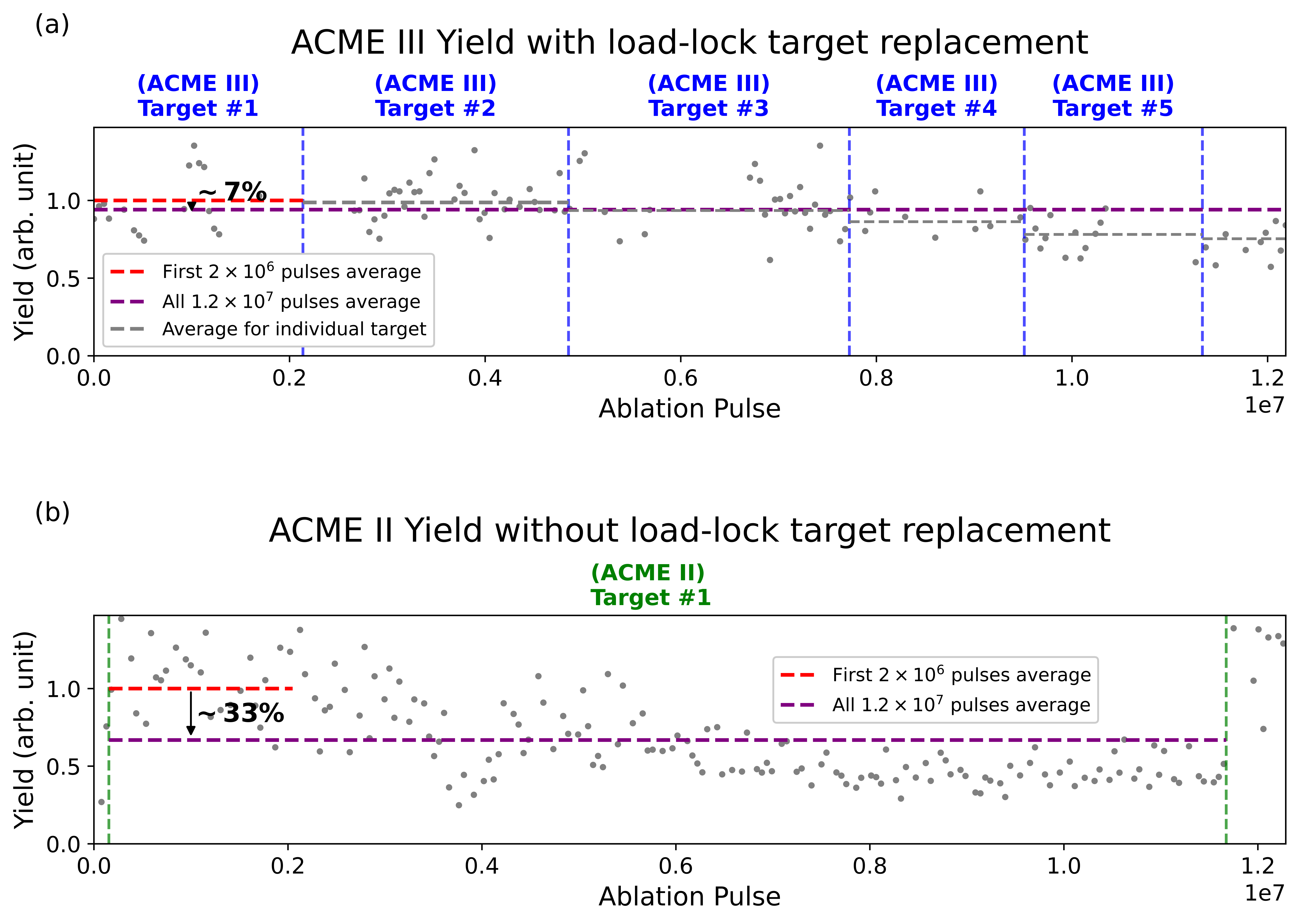}
  \caption{Yield in (a) ACME III ongoing operation with load-lock target replacement and (b) part of the ACME II final run without load-lock target replacement. Each gray scatter dot represents the yield averaged over $5\times10^4$ consecutive pulses. The yield is normalized to the average yield of the first $2\times10^6$ pulses. Gaps in scattering plots are from exclusion of data taken under conditions very different from eEDM measurement for systematic error checking purposes. Blue vertical dashed lines in (a) indicate a target replacement with the load-lock system without cleaning the dust in cell, and green vertical dashed lines in (b) indicate a target replacement by hand and with cleaning the dust in cell. The full ACME II eEDM final run consists of $\sim4.3\times10^7$ pulses. }
  \label{fig:longtermyield}
\end{figure*}

\subsection{Beam Properties}

We characterize several properties of the beam, both immediately after installing the target plate with the load-lock system and installing through the old, standard procedure, where the system is warmed up and beam box vacuum is broken \cite{hutzler2011cryogenic}. All measurements are performed with a neon buffer gas flow rate of 40~standard cubic centimeters per minute (sccm) and a cell temperature of \SI{18}{\kelvin}.

The number of molecules produced per ablation pulse is estimated from the integral of the optical density of the beam right outside of the cell aperture over time as \cite{wright2023cryogenic}:
\begin{equation}
        N=\frac{v_{\parallel}A_\mathrm{aperture}}{n_{\mathrm{pass}}l\sigma_D} \int{OD\left( t \right) dt},
\end{equation}
where $v_{\parallel}$ is the forward velocity of beam, $n_{\mathrm{pass}}=2$ is the number of passes of the absorption laser, and $l$ is assumed to be the aperture diameter $\sim$\SI{5}{\milli\meter}. The absorption cross section at the peak of the Doppler profile $\sigma_D$ is given by $
    \sigma _D=\frac{\sqrt{\pi}}{2}\frac{\gamma}{\Gamma _D}\sigma _0$,
where $\Gamma_D$ is the Doppler width measured right outside of the cell and $\gamma$ is the natural linewidth of the excited state $C$. The resonant absorption cross section $\sigma_0$ is estimated from the measured radiative decay lifetime \cite{kokkin2014branching} and the calculated Franck-Condon factors \cite{wentink1972isoelectronic} of the $C$ state.
With the absorption laser driving the Q(1) branch of the $C(v^{\prime}=0)\leftarrow X(v=0)$ transition, we measure $\sim3.8\times10^{11}$ molecules in the $J=1$ state in the beam per ablation pulse. At a rotational temperature of 4.8~K, we infer that $\sim1.3\times10^{11}$ molecules in the ground state $J=0$ are produced with each ablation pulse, which is close to the $\sim1\times10^{11}$ ground state molecules per pulse achieved by a cell with similar condition and geometry but with a target plate installed under the old process \cite{hutzler2011cryogenic}. The extraction fraction, which is given by the density ratio right outside of the cell aperture to that inside of the cell, is $\epsilon=11\%$ as shown in Fig.\ref{fig:beam_properties_combined}~(a); this is also consistent with the old process \cite{hutzler2011cryogenic}.

The rotational temperature of the molecules in the beam is measured by absorption spectroscopy performed 2.5~cm downstream of the cell aperture (before the beam collimator). The rotational population $p_J$ for $J=0,1,2,3$ is measured by scanning the absorption laser around the R(0), Q(1), Q(2) and Q(3) branches of the $C(v^{\prime}=0)\leftarrow X(v=0)$ transition. Fig.\ref{fig:beam_properties_combined}~(b) shows the result. Assuming a Boltzmann distribution, the rotational temperature $T$ is fitted from 
\begin{equation}
\frac{p_J}{p_0}=\left( 2J+1 \right) e^{-\frac{B_eJ\left( J+1 \right)}{k_BT}},
\end{equation}
where $B_e=0.33264$~cm$^{-1}$ is the rotational constant of the $X$ state \cite{hutzler2014new,dewberry2007pure}. With a load-lock-installed ThO$_2$ target plate, the beam is measured to have a rotational temperature of \SI{4.8 \pm 0.7}{\kelvin}, which is consistent with the value of \SI{4.6}{\kelvin} measured with the old process \cite{hutzler2011cryogenic}.

The same rotational absorption spectra also provide the FWHM of the transverse velocity $\Delta v_{\perp}$ of the beam at \SI{2.5}{\centi\meter} downstream, and the divergence angle is \cite{hutzler2012buffer} $
\theta _{\mathrm{FWHM}}=2\mathrm{arctan} \left( \frac{\Delta v_{\bot}/2}{v_{\parallel}} \right)$. We have $\theta_{\mathrm{FWHM}}=36\degree$, corresponding to a half-maximum solid angle $\Omega=2\pi(1-\cos(\theta_{\mathrm{FWHM}}/2))=$ \SI{0.31}{\steradian} and close to the hydrodynamic entrainment limit of $\pi m_{\mathrm{Ne}}/m_{\mathrm{ThO}}\approx0.3$, similar to the value measured with old installation process \cite{hutzler2011cryogenic}.

The forward velocity distribution of the beam is shown in Fig.\ref{fig:beam_properties_combined}~(c). The forward velocity is extracted from the Zeeman phase component of the spin precession phase measured \(\sim\)\SI{2}{\meter} downstream (Fig.\ref{beambox}~(c)). With a clean cell (where there is no build up of ablation dust), we measure the forward velocity of the beam $v_{\parallel}$ to be \(\sim\)\SI{200}{\meter\per\second}, consistent with the result measured from the old process \cite{hutzler2011cryogenic}. This forward velocity is close to $\sqrt{5k_BT_0/m_{\mathrm{Ne}}}\approx$ \SI{193}{\meter\per\second}, the final velocity of a supersonic expansion of the monoatomic carrier gas neon at \SI{18}{\kelvin} (aka. `fully boosted' \cite{hutzler2012buffer}). After four targets have been ablated, depleted and replaced with the load-lock system, but without any cleaning of the dust in the cell, we again measure the forward velocity with the $5^{\mathrm{th}}$ target and observe \(\sim\)\SI{215}{\meter\per\second}.

We speculate that the increase in forward velocity originates from the thermalization process of the buffer gas \cite{hutzler2012buffer,skoff2011diffusion} being affected by dust accumulated from ablation. The mass of ablation dust accumulated after four targets is estimated to be a few grams, and the interior of the cell is seen to be fully covered by dust. The dust could reduce the thermal conductivity between the cell and the buffer gas neon, causing a higher instantaneous neon temperature after ablation and thus a faster beam. The phenomenon of dustier cells producing faster beams has also been observed in other experiments using CBGBs \cite{mooij2024influence,wright2023cryogenic, white2024slow}.

\subsection{Long-term yield improvement}

We characterize the long-term yield improvement enabled by the load-lock target replacement system by comparing the yield data from the final run of the ACME II experiment \cite{acme2018improved} (Fig.\ref{fig:longtermyield}~(a)) and the ongoing operation of the ACME III experiment (Fig.\ref{fig:longtermyield}~(b)). State transfer and detection efficiencies are different between the two experiments, but the number of fluorescent photoelectrons detected are proportional to the molecular yield from the buffer gas cell for both experiments. The buffer gas cells in ACME II and ACME III have similar geometries and baseline yields, with the difference being that the load-lock target replacement system is implemented only in ACME III.

In ACME II, a target is usually ablated for $1.2\times10^7$ laser pulses over a one-month period before the cleaning of the cell and the replacement of the target take place. The yield declines significantly during the whole period. After the first  \(2 \times 10^6\) pulses, the fresh surface area is depleted, and the yield drops by $\sim50\%$. While adjusting the focal point of the ablation laser could provide a temporary improvement, it could not restore stable yields once an uneven, porous ablation surface has formed. Assuming the average yield per shot during the initial \(2 \times 10^6\) pulses---corresponding to a fresh target and clean cell---represents 100\%, the average yield per shot over the entire \(1.2 \times 10^7\) pulses is  $\sim67\%$ of this initial value.

In ACME III, the load-lock target replacement system allows in-situ replacement of the target once the fresh surface area is depleted. We start with a clean cell and six targets are used over \(1.2 \times 10^7\) pulses. No significant degradation of the yield is observed until the third target and, by the sixth target, the yield decreases by \(\sim25\%\) compared to the first, presumably due to dust accumulation in the cell. We note that the rate of dust accumulation is not expected, nor observed, to be affected by the use of the load-lock system. Over the whole period of $1.2\times10^7$ pulses before cell dust cleaning takes place, with target replacement done regularly with the load lock system, the average yield is $\sim93\%$ of the average yield during the initial \(2 \times 10^6\) pulses. Compared to the case without target replacement with the load-lock system, the long-term average yield is improved by $\sim40\%$. Overall, the load-lock system maintains higher molecular flux, reduces the need for vacuum venting and thermal cycling of the CBGB, and enables more continuous operation of ACME III.

\section{Conclusion}

We design, implement and characterize a load-lock CBGB used in the ACME III experiment. This source enables in situ replacement of ThO$_2$ ablation targets without interrupting vacuum or cryogenic conditions. 
With a target assembly installed via the load-lock system, this CBGB produces $1.3\times10^{11}$ ground state molecules per ablation pulse at a cell extraction fraction of $11\%$, a rotational temperature of \SI{4.8}{\kelvin} and a half-maximum solid angle of \SI{0.31}{\steradian}, all of which are consistent with results from a buffer gas cell with a target plate installed by warming the entire CBGB and breaking vacuum. We observe an increase in beam forward velocity from \SI{200}{\meter\per\second} to \SI{215}{\meter\per\second} due to dust accumulation in the cell over the course of using four targets.

We also characterize the long-term yield improvement enabled by the load-lock system with the yield history by comparing the ACME III systematic error checking run and ACME II final run. By regularly replacing targets when the fresh surface for ablation is depleted, the long term yield is improved by $\sim40\%$. We expect the load-lock system to continue maintaining high molecular flux for the ACME III experiment. Although developed for laser ablation, the lock-lock system is broadly applicable to other CBGB sources requiring in-situ precursor replacement under vacuum or cryogenic conditions.

\section{acknowledgments}

We would like to thank N. Hutzler, C. Panda, B. Augenbraun and Y.Bao for discussions and insights. We also acknowledge the contribution of S. Rodriguez, T. Bui, S. Cotreau, and S. Sansone. This material is based upon work supported by the National Science Foundation, the Gordon and Betty Moore Foundation, the Alfred P. Sloan Foundation, JSPS Kakenhi, and Okayama University RECTOR program.

\section*{Data Availability}

The data that support the findings of this study are available from the corresponding author upon reasonable request.

\bibliographystyle{prsty_gg}
\bibliography{NewRefs}

\end{document}